\documentclass[twocolumn]{aastex6}
\usepackage{natbib}
\usepackage{pslatex}
\usepackage{graphicx}
\newcommand{\Htwoo}{\mbox{H$_{2}$O}}

\newcommand{\Ntwo}{\mbox{N$_{2}$}}
\newcommand{\cotwo}{\mbox{CO$_{2}$}}
\newcommand{\otwo}{\mbox{O$_{2}$}}

\newcommand{\kms}{km~s$^{-1}$}

\newcommand{\lyb}{{Lyman-$\beta$}}

\shorttitle{Gas Outbursts of Comet 67P}
\shortauthors{Feldman, et al.}

\begin{document}

\title{The Nature and Frequency of the Gas Outbursts in Comet 67P/Churyumov-Gerasimenko observed by the Alice Far-ultraviolet Spectrograph on Rosetta} 

\author{Paul D. Feldman}
\affil{Department of Physics and Astronomy, The Johns Hopkins University, 3400 N. Charles Street, Baltimore, Maryland 21218}
\email{pfeldman@jhu.edu}
 
\author{Michael F. A'Hearn and Lori M. Feaga}
\affil{Astronomy Department, University of Maryland, College Park, MD 20742}

\author{Jean-Loup Bertaux}
\affil{LATMOS, CNRS/UVSQ/IPSL, 11 Boulevard d'Alembert, 78280 Guyancourt, France}

\author{John Noonan, Joel Wm. Parker, Eric Schindhelm, Andrew J. Steffl and S. Alan Stern}
\affil{Southwest Research Institute, Department of Space Studies, Suite 300, 1050 Walnut Street, Boulder, CO 80302}

\and

\author{Harold A. Weaver}
\affil{Space Exploration Sector, Johns Hopkins University Applied Physics Laboratory, 11100 Johns Hopkins Road, Laurel, MD 20723-6099}

\pagestyle{myheadings}

\begin{abstract}

Alice is a far-ultraviolet imaging spectrograph onboard {\it Rosetta} that, amongst multiple objectives, is designed to observe emissions from various atomic and molecular species from within the coma of comet 67P/Churyumov-Gerasimenko.  The initial observations, made following orbit insertion in August 2014, showed emissions of atomic hydrogen and oxygen spatially localized close to the nucleus and attributed to photoelectron impact dissociation of \Htwoo\ vapor.  Weaker emissions from atomic carbon were subsequently detected and also attributed to electron impact dissociation, of \cotwo, the relative \ion{H}{1} and \ion{C}{1} line intensities reflecting the variation of \cotwo\ to \Htwoo\ column abundance along the line-of-sight through the coma.  Beginning in mid-April 2015, Alice sporadically observed a number of outbursts above the sunward limb characterized by sudden increases in the atomic emissions, particularly the semi-forbidden \ion{O}{1} $\lambda$1356 multiplet, over a period of 10-30 minutes, without a corresponding enhancement in long wavelength solar reflected light characteristic of dust production.  A large increase in the brightness ratio \ion{O}{1} $\lambda$1356/\ion{O}{1} $\lambda$1304 suggests \otwo\ as the principal source of the additional gas.  These outbursts do not correlate with any of the visible images of outbursts taken with either OSIRIS or the navigation camera.  Beginning in June 2015 the nature of the Alice spectrum changed considerably with CO Fourth Positive band emission observed continuously, varying with pointing but otherwise fairly constant in time.  However, CO does not appear to be a major driver of any of the observed outbursts.

\end{abstract}

\keywords{comets: individual (67P) --- ultraviolet: planetary systems}

\newpage
\section{INTRODUCTION}

We have previously \citep{Feldman:2015} described the initial observations of the near-nucleus coma of comet 67P/Churyumov-Gerasimenko made by the Alice far-ultraviolet imaging spectrograph onboard {\it Rosetta} made in the first few months following orbit insertion in August 2014. These observations of the sunward limb, made from distances between 10 and 30 km from the comet's nucleus, showed emissions of atomic hydrogen, oxygen, and carbon, spatially localized close to the nucleus and attributed to photoelectron impact dissociation of \Htwoo\ and \cotwo\ vapor.  This interpretation is supported by measurements of suprathermal electrons by the Ion and Electron Sensor (IES) instrument on {\it Rosetta} \citep{Clark:2015}. Beginning in February 2015, as the activity of the comet increased, the orbit of {\it Rosetta} was adjusted to increasing distance from the nucleus due to concerns for spacecraft safety.  At distances $\geq$50~km, when pointed towards the nadir, the spatial extent along the Alice 5.5\arcdeg\ long slit \citep{Stern:2007} allows the coma, both sunward and anti-sunward, to be resolved from the nucleus and observed nearly continuously.  

This geometry allowed Alice, beginning in mid-April 2015, to observe a number of outbursts above the sunward limb.  These outbursts are characterized by sudden increases in the atomic emissions, particularly the semi-forbidden \ion{O}{1} $\lambda$1356 multiplet, over a period of 10-30 minutes, without a corresponding enhancement in long wavelength solar reflected light characteristic of dust production.  The corresponding increase in the brightness ratio \ion{O}{1} $\lambda$1356/\ion{O}{1} $\lambda$1304 suggests that \otwo, detected for the first time in a comet by \citet{Bieler:2015}, is the primary source of the additional gas.  
This is the same spectroscopic diagnostic used to determine that \otwo\ is the dominant species in the exospheres of Europa and Ganymede \citep{Hall:1995,Hall:1998}.  As the comet rotates the Alice slit samples different regions above the comet's limb, but the magnitude of the increase on the short time scale as well as the spatial distribution along the slit makes it very unlikely that it is due to a spatial gradient or a collimated ``jet''.

Although \ion{C}{1} $\lambda$1657 is also seen to increase, the variation in the brightness ratio \ion{O}{1} $\lambda$1356/\ion{C}{1} $\lambda$1657 indicates that it cannot be \cotwo\ alone, nor can the effects be due to an increase in photoelectron flux.  These outbursts do not correlate with any of the visible images of outbursts taken with either OSIRIS \citep{lin:2016,Vincent:2016} or the navigation camera.  We also find that CO does not appear to be driving any of the observed outbursts.

\section{OBSERVATIONS}

\subsection{Instrument Description}

Alice is a lightweight, low-power, imaging spectrograph designed for in situ far-ultraviolet imaging spectroscopy of comet 67P in the spectral range 700-2050~\AA.  The slit is in the shape of a dog bone, 5.5\degr\ long, with a width of 0.05\degr\ in the central 2.0\degr, while the ends are 0.10\degr\ wide, giving a spectral resolution between 8 and 12~\AA\ for extended sources that fill its field-of-view.  Each spatial pixel or row along the slit is 0.30\degr\ long.  Details of the instrument have been given by \citet{Stern:2007}.

\subsection{Gas Outbursts}

\begin{figure}[ht]
\begin{center}
\includegraphics*[width=0.5\textwidth,angle=0.]{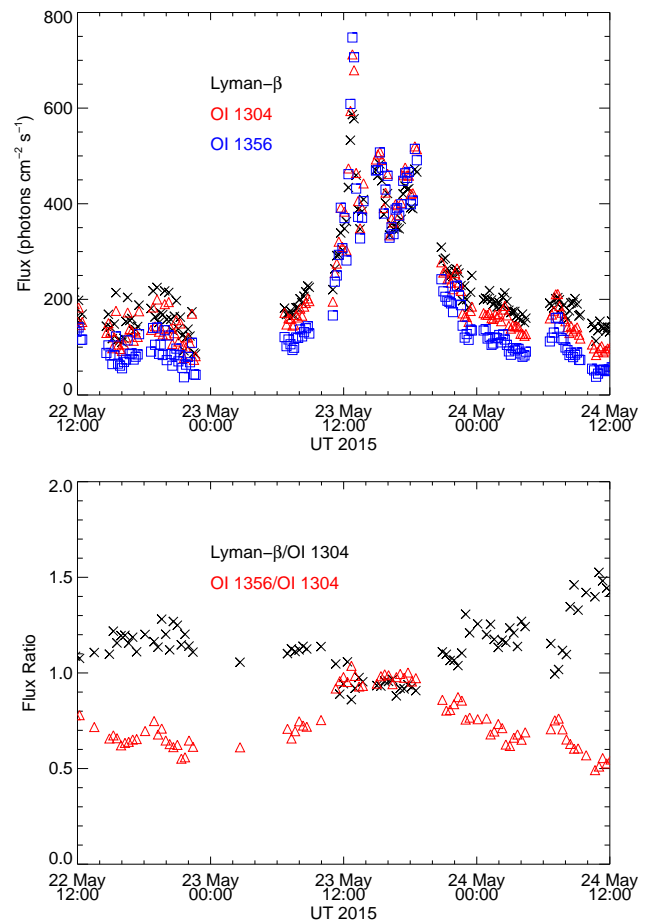}
\caption {Temporal variation of the atomic hydrogen and oxygen emissions above the limb on 2015 May 22--24 (top). Individual error bars are not shown but the 1-$\sigma$ statistical uncertainty for all points is $<$5\%.  The bottom panel shows the variation in relative intensities of these emissions.  \label{light_curve}}
\end{center}
\end{figure}

The strongest of the outbursts, from the period May-July 2015, are listed in Table~\ref{gas}.  The table includes the heliocentric distance of the comet, $r_h$, the distance of {\it Rosetta} from the center of the comet, $d$, the sub-spacecraft longitude and latitude, and the solar phase angle at the time of observation.  With the exception of the June 18 and June 23 outbursts, the sub-spacecraft positions of the tabulated outburst are uncorrelated, attesting to their random nature.  As an example we will consider in detail the outburst of May 23 since the pointing was constant over two rotations of the comet enabling us to obtain nearly continuous light curves except for gaps in the data due to a lack of observations during spacecraft maintenance activities.  Light curves for the strongest coma emissions are shown in Fig.~\ref{light_curve}.  Note that prior to the outburst the relative intensities of \ion{H}{1} \lyb, \ion{O}{1} $\lambda$1304, and \ion{O}{1} $\lambda$1356, are consistent with the earlier observations that attributed the emissions primarily to electron dissociative excitation of \Htwoo\  \citep{Feldman:2015}.  One rotation period ($\sim$12~h) later both the absolute and relative brightnesses have returned to their pre-outburst values, \deleted{suggesting that the source of volatile ice driving the outburst is located deeper than the skin depth of the diurnal heat wave.}\added{in strong contrast to the persistent diurnal variability seen in mass spectrometer data \citep{Hassig:2015,Mall:2016,Fougere:2016}.}

\begin{figure*}[ht]
\begin{center}
\includegraphics*[width=0.37\textwidth,angle=0.]{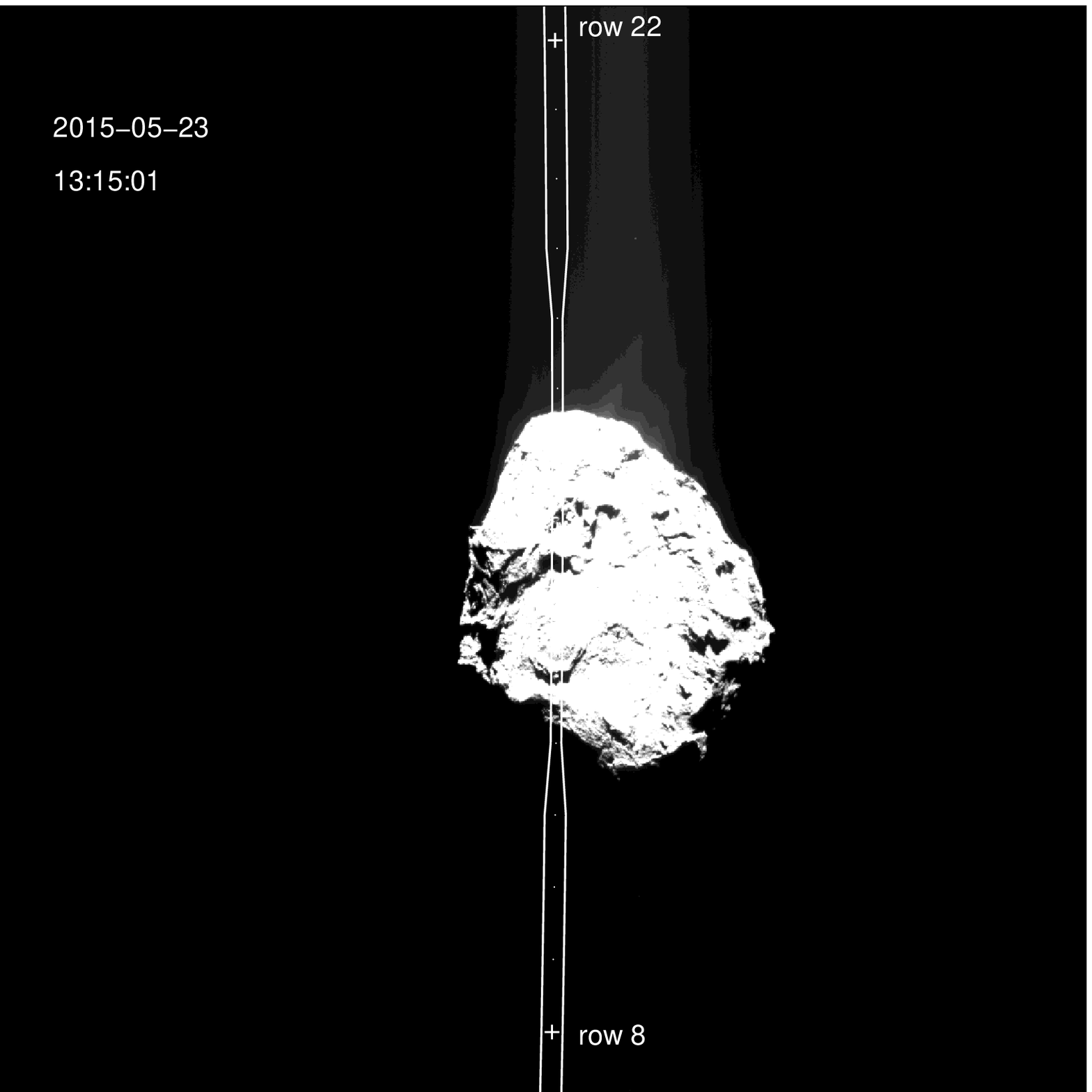}
\includegraphics*[width=0.62\textwidth,angle=0.]{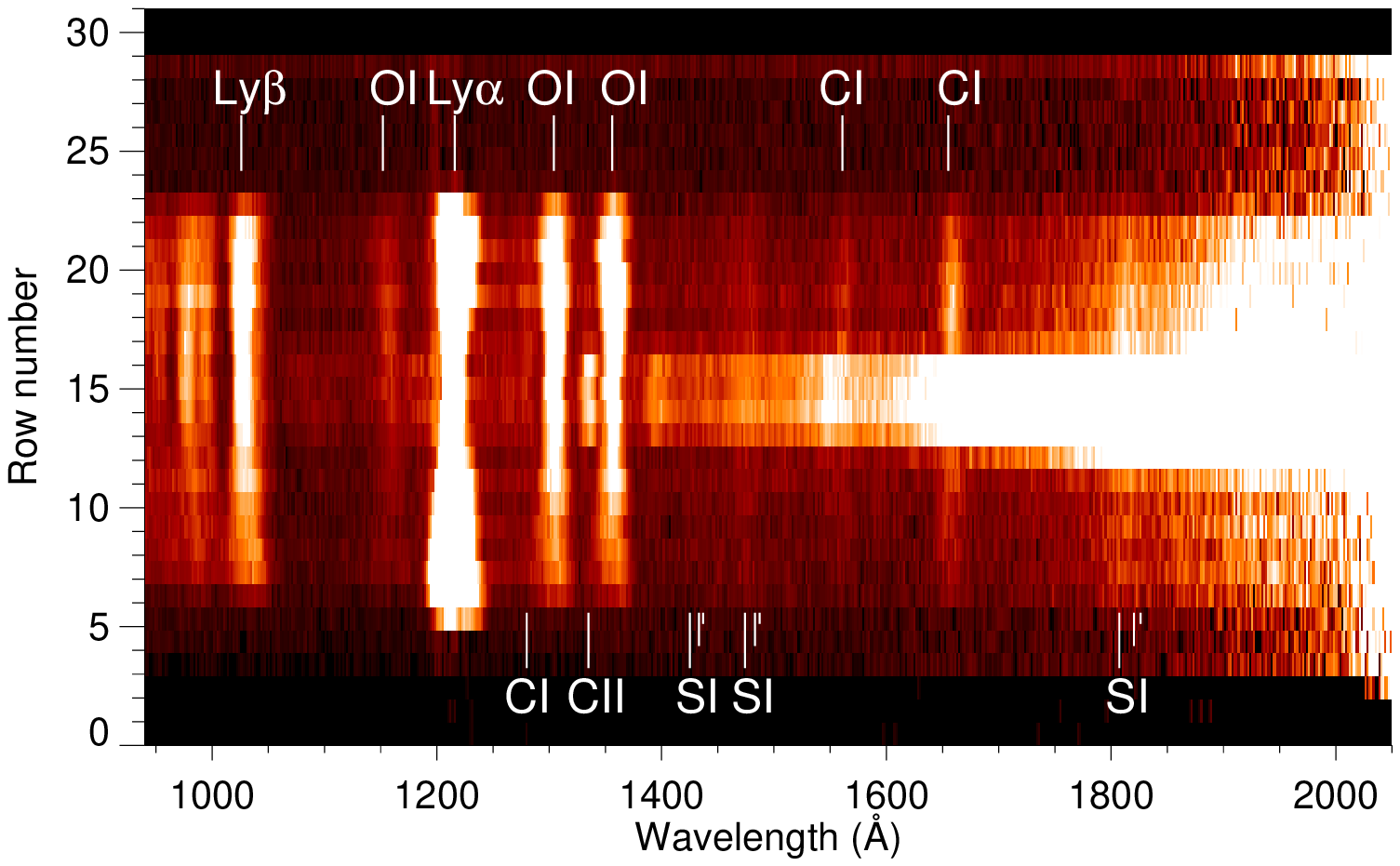}
\caption{Left: NAVCAM context image obtained 2015 May 23 UT 13:15, shortly after the peak emission was observed.  Right: Spectral image beginning UT 12:31, 1589 s exposure, three co-added histograms from the time of peak emission.  The sunlit nucleus appears in rows 12--17.  The distance to the comet was 143 km, the heliocentric distance was 1.58~AU, and the solar phase angle was 61.1\degr.  For both images the direction of the Sun is towards the top of the image.  \label{nav1} }
\end{center}
\end{figure*}

The orientation of the Alice slit projected onto the comet is shown in Fig.~\ref{nav1} (left) in an image from the navigation camera (NAVCAM) taken $\sim$45 minutes after the peak brightness.  The Sun is towards the top of the image  \added{and the comet's rotation axis is roughly perpendicular to the slit.  We attribute the secondary peaks seen in the light curve (top panel of Fig.~\ref{light_curve}) to geometric effects of the rotation, as visualized in the three-dimensional coma models of \citet{Fougere:2016}.  Similar effects are seen in the light curves of the other outbursts listed in Table~\ref{gas}}.  

The spectral image in Fig.~\ref{nav1} (right) shows the clearly separated coma emissions together with the reflected solar spectrum from the nucleus in rows 11 to 17 of the slit.  Again, the Sun is towards the top of the image.  Note that the \ion{O}{1} emissions are seen against the solar reflected radiation from the nucleus and into the anti-sunward coma.

\begin{deluxetable*}{lccccccc}
\tabletypesize{\small}
\tablewidth{0pt}
\tablecolumns{9}
\tablecaption{Major Gaseous Outbursts Observed by Alice in May - July 2015. \label{gas}}
\tablehead{
\colhead{Date} & \colhead{Peak} & \colhead{$r_h$} & \colhead{$d$} & \multicolumn{2}{c}{Sub-spacecraft} &  \colhead{Phase} & \colhead{$B_{max}$(1356)} \\
\colhead{} & \colhead{Time (UT)\tablenotemark{a}} & \colhead{(AU)} & \colhead{(km)} & \colhead{Longitude (\arcdeg)} &  \colhead{Latitude (\arcdeg)} &  \colhead{Angle (\arcdeg)} & \colhead{(rayleighs)\tablenotemark{b}} }
\startdata
2015 May 23 & 12:42 & 1.58 & 143 & 151.2 & --17.1 &  61.1 & 223 \\
2015 June 18 & 03:43 & 1.42 & 202 & 214.8 & 51.3 &  89.9 & 173 \\
2015 June 20 & 15:26 & 1.40 & 181 & 269.3 & 19.9 &  89.9 & 60 \\
2015 June 23 & 20:39 & 1.39 & 196 & 215.0 & 53.3 &  89.7 & 113 \\
2015 July 04 & 08:56 & 1.34 & 179 & 111.6 & 53.8 &  89.9 & 83 \\
2015 July 13 & 01:16 & 1.30 & 155 & 149.3 & 15.6 &  88.8 & 78 \\
\enddata
\vspace*{-2pt}
\tablenotetext{a}{start time of histogram integration}
\vspace*{-2pt}
\tablenotetext{b}{maximum \ion{O}{1} $\lambda$1356 brightness above the sunward limb in a 0.3\arcdeg\ spatial pixel.}
\end{deluxetable*}

\subsection{Spectra}

\begin{figure*}[ht]
\begin{center}
\includegraphics*[width=0.45\textwidth,angle=0.]{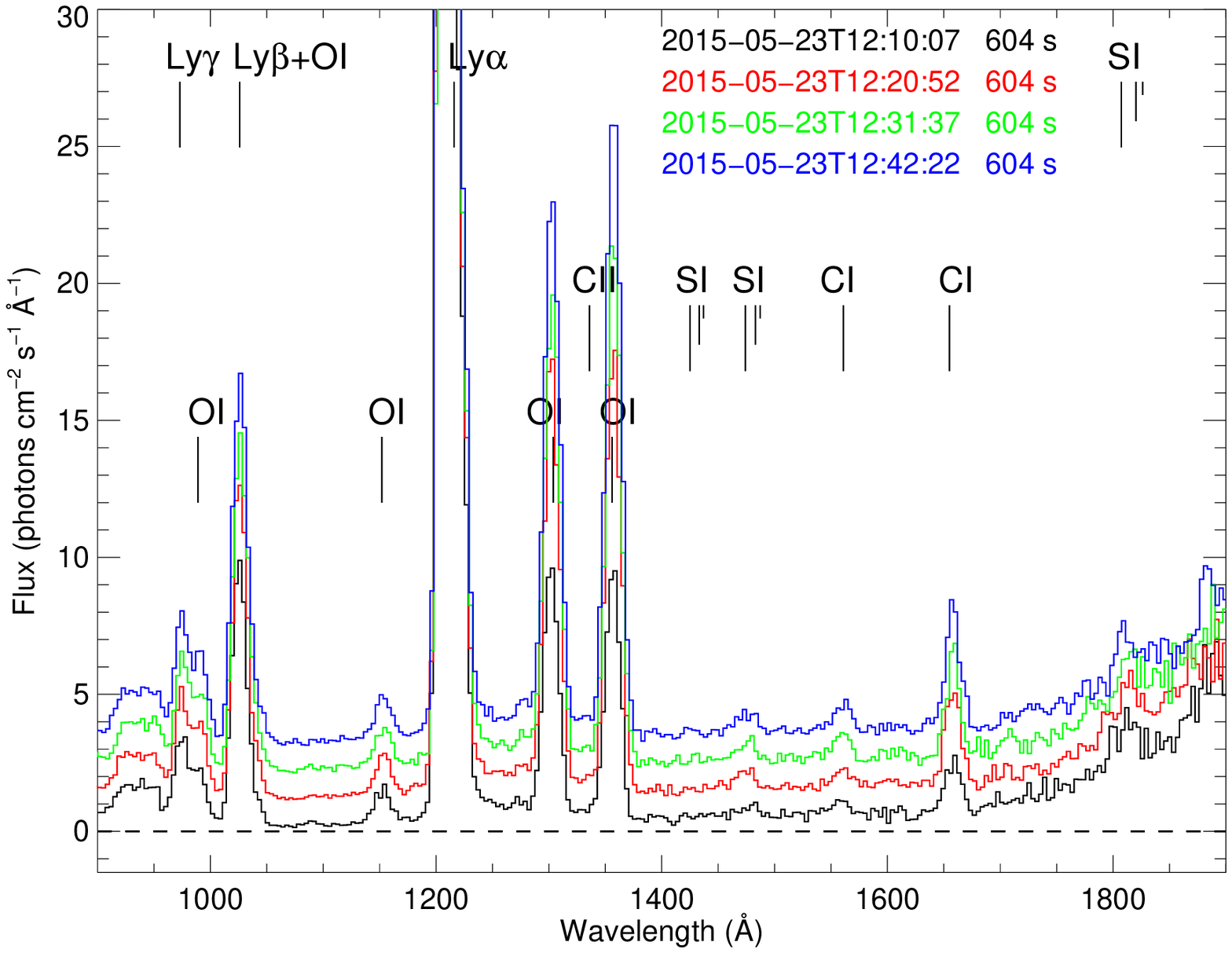}
\includegraphics*[width=0.45\textwidth,angle=0.]{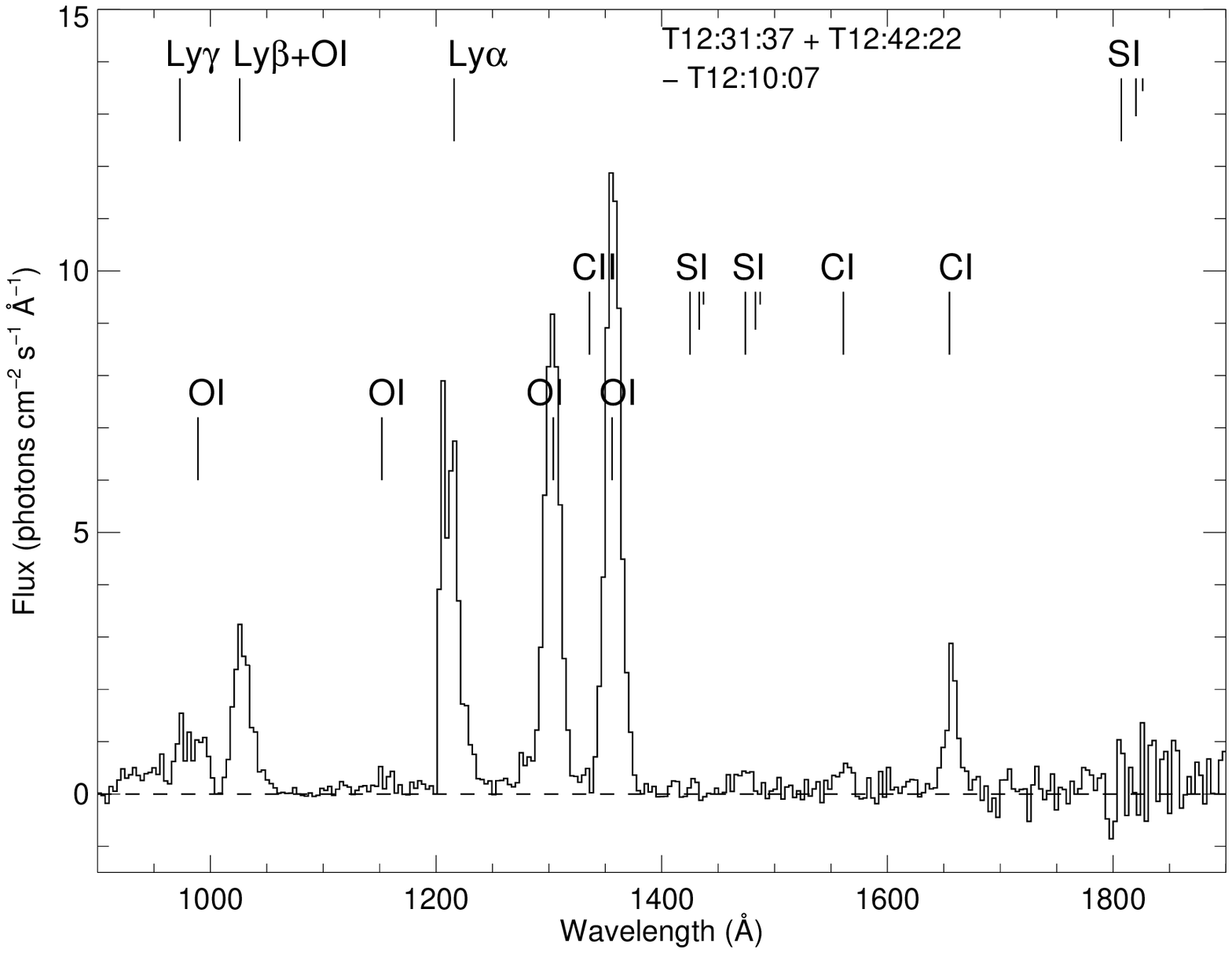}
\caption {Left: Sequence of four 10-minute histogram spectra above the sunlit limb.  The spectra are offset from one another for clarity.  The upturn at long wavelengths is due to scattered solar radiation from dust in the coma.  Right: The difference between the average of the final two and the first of the spectra shown at left.  The difference represents the spectrum of the material ejected into the coma in a $\sim$30 minute period.  \label{may23_4spec}}
\end{center}
\end{figure*}

To study the evolution of the gas content during the outburst we present four successive spectra of the sunward coma taken from 10-minute histograms beginning 2015 May 23 UT12:10:07 in the left panel of Fig.~\ref{may23_4spec}.  These correspond to rows 18 to 21 of the spectral image in Fig~\ref{nav1}.  The difference between the spectra at the peak of the outburst and the prior spectra is then indicative of the erupting gas.
The difference spectrum, shown in the right panel, is characterized by an enhancement in the atomic emissions, particularly atomic oxygen, without a corresponding enhancement in long wavelength solar reflected light characteristic of dust production.  The large increase in \ion{O}{1} $\lambda$1356/\ion{O}{1} $\lambda$1304 (intensity ratio $\geq$1:1) suggests \otwo, now known to be present in the cometary ice of 67P \citep{Bieler:2015}, as the primary source of the additional gas. 

However, laboratory data on the electron impact dissociative excitation of \otwo\ \citep{Kanik:2003} suggests that this ratio should be $\sim$2, as seen in the exospheric spectra of Europa \citep{Hall:1995} and Ganymede \citep{Hall:1998}.  The observed ratio varies in the other outbursts listed in Table~\ref{gas} and likely reflects additional sources of \ion{O}{1} $\lambda$1304 such as photodissociation of \otwo\  \citep{Beyer:1969,Lee:1974} or electron impact on O (if present).
The dramatic change in relative intensities implies that the outburst cannot be due to a sudden increase in photoelectron flux or change in the electron energy distribution. For May 23 this is borne out by data from RPC/IES \citep{Clark:2015} that shows only a modest uniform increase in electron flux beginning about two hours before the outburst observed by Alice and lasting for 12 hours (K. Mandt, private communication).

A significant amount of \Htwoo\ is also released as evidenced by the presence of \ion{H}{1} Lyman-$\alpha$ and Lyman-$\beta$ in the difference spectrum.  If we ignore the blending of Lyman-$\beta$ with \ion{O}{1} $\lambda$1027, also produced by e+\otwo\ \citep{Ajello:1985}, and assume that all of the emission at 1026~\AA\ is Lyman-$\beta$, then we can use the e+\Htwoo\ cross section for Lyman-$\beta$ at 100~eV from \citet{Makarov:2004} relative to the e+\otwo\ cross section for \ion{O}{1} $\lambda$1356 from \citet{Kanik:2003} to estimate the relative \otwo /\Htwoo\ abundance in the outburst.  From the relative fluxes in the difference spectrum (Fig.~\ref{may23_4spec}) we find \otwo /\Htwoo\ $\geq$~0.5, considerably greater than the mean quiescent value of 0.038 reported by \citet{Bieler:2015}.
The presence of \Htwoo\ in the outburst is confirmed by concurrent sub-mm measurements of the \Htwoo\ column density by MIRO \citep{Lee:2015} along a line-of-sight contiguous with the central row of the Alice slit that showed a threefold increase at the same time (P. von Allmen, private communication), \added{consistent with the increase in Lyman-$\beta$ seen in the top panel of Fig.~\ref{light_curve}}. 

CO began to appear regularly in Alice coma spectra in June 2015 and continues to be present through early 2016.  While not detected in the May 23 spectrum the CO Fourth Positive system is seen in several other spectra listed in Table~\ref{gas}.  However, it does not appear in any of the difference spectra and thus is unlikely to be the major driver in any of the outbursts presented here.
The origin of the \ion{C}{1} $\lambda$1657 emission in the difference spectrum is not clear.  Electron impact on C\otwo\ would produce emission at 1561 \AA\ with an intensity about half that of \ion{C}{1} $\lambda$1657 \citep{Mumma:1972} as well as CO Cameron band emission at longer wavelengths which is not observed.  Other carbon bearing molecules are not precluded as a source \added{\citep{LeRoy:2015}}.

\added{We note that \Ntwo, also discovered for the first time in 67P \citep{Rubin:2015}, also has a rich electron excited spectrum of \Ntwo\ bands and atomic and ionic N multiplets within the Alice spectral range \citep{Heays:2014}.  None of these are detected in the outburst spectrum so that \Ntwo, whose mean abundance relative to CO is found by \citeauthor{Rubin:2015} to be less than 1\%, also plays no role in the gas outbursts.}

\subsection{Spatial Profiles \label{spatial}}

\begin{figure*}[ht]
\begin{center}
\includegraphics*[width=0.48\textwidth,angle=0.]{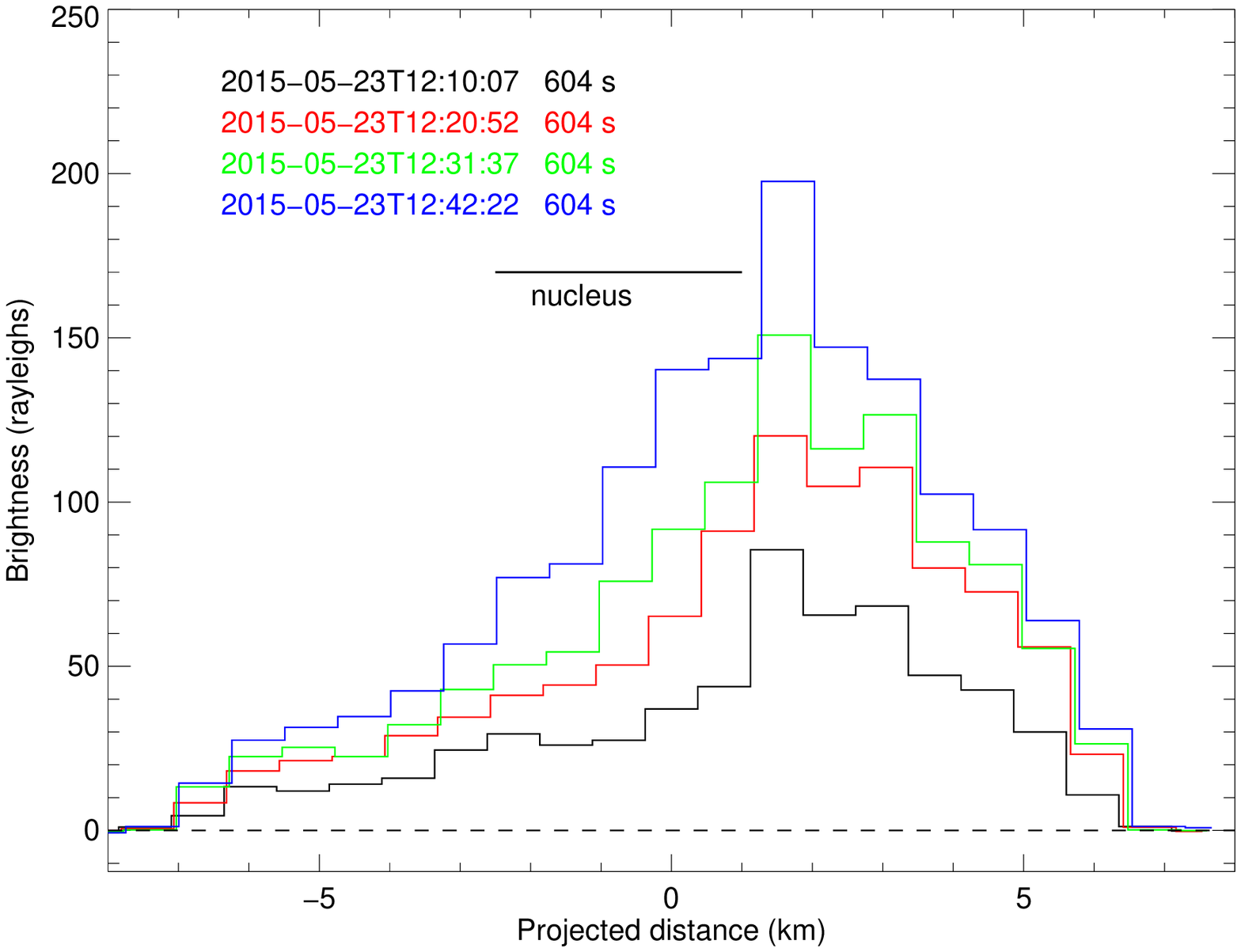}
\includegraphics*[width=0.48\textwidth,angle=0.]{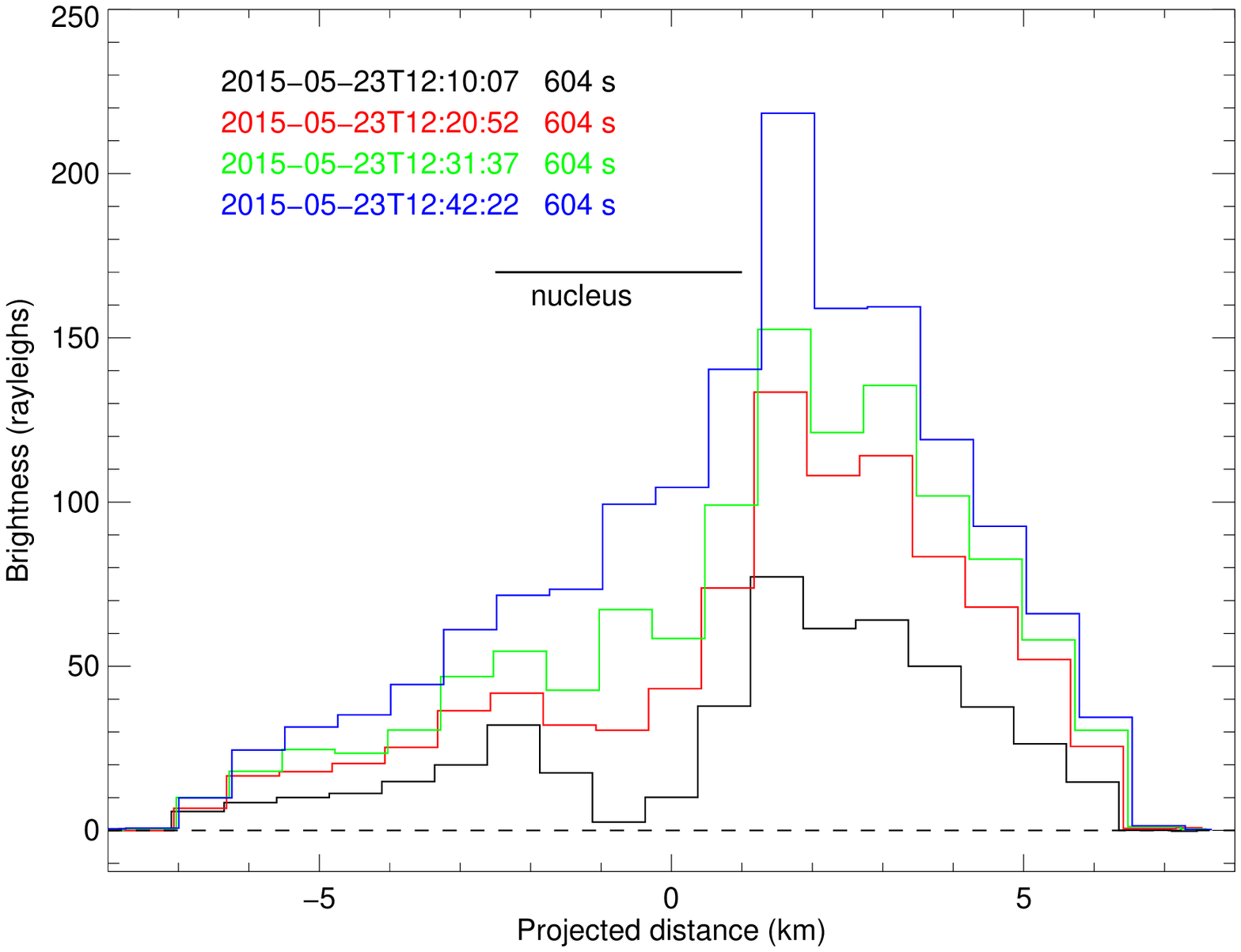}
\caption {Spatial profiles of \ion{O}{1} $\lambda$1304 (left) and \ion{O}{1} $\lambda$1356 (right) corresponding to the spectra shown in Fig.~\ref{may23_4spec}.  The Sun is to the right.  The position of the nucleus is indicated.  The ends of the projected slit are at --6.5 and +6.0~km, respectively.  \label{may23_spatial}}
\end{center}
\end{figure*}

Additional information about the outburst can be obtained from the profiles of the \ion{O}{1} emissions along the slit.  Fig.~\ref{may23_spatial} shows the profiles of \ion{O}{1} $\lambda$1304 and \ion{O}{1} $\lambda$1356 for the four spectra shown in Fig.~\ref{may23_4spec}.  Both profiles peak just above the sunward limb and decrease radially outward.  The brightness increases nearly uniformly in each successive 10 minute integration.  
With an outflow velocity of 0.5 \kms\ the escaping \otwo\ molecules will exit the Alice field-of-view in $\sim$10 seconds, so the ejection is continuous.  From the light curve we see that one rotation later ($\sim$12 h), at the same sub-spacecraft longitude, the emission has returned to its quiescent level.
Emission is seen against the nucleus and off the anti-sunward limb, indicating that the outburst is not in the form of a collimated jet but is rather diffuse.  As noted above, any dust produced would have been detected by an increase at long wavelengths of reflected solar radiation.

\subsection{Additional Events \label{add}}

For the other dates in Table~\ref{gas}, the light curves, difference spectra, and spatial profiles are all similar to those of the May 23 event.
A search through the Alice database for earlier events reveals multiple outbursts with similar spectra on 2015 April 15 and April 29/30.  Prior to April, the geometry for observing outbursts (closer distance to the nucleus) was less favorable.  In April the spacecraft was at southern latitudes but the longitudes of the outbursts was also random.  The rate of detected gas outburst events decreased after perihelion on 2015 August 13.
Post-perihelion observations, through the present, continue to show variable \ion{O}{1} $\lambda$1356/\ion{O}{1} $\lambda$1304 intensity ratios although these measurements are not always confined to the short time scales of the ``outburst'' events described above.
These observations provide a means of monitoring the \otwo/\Htwoo\ abundance in the coma even when {\it Rosetta} is at large distances ($\geq$100 km) from the nucleus, complementing ROSINA measurements closer to the comet.

\section{DISCUSSION}

A likely source of gas outbursts is the warming of sub-surface volatile reservoirs as the comet approaches perihelion.  Water ice containing frozen \otwo, as has been proposed for the surface of Ganymede \citep{Spencer:1995}, would reside below the dust mantle. Sublimated \otwo, together with some \Htwoo, would then percolate through the porous mantle and diffuse into the coma taking some of the dust with it.
The coupling of dust to gas sublimated from sub-surface ice has been studied by many authors \citep[e.g.,][]{Blum:2014}  For 67P, \citet{Gundlach:2015} consider only CO, \cotwo, and \Htwoo\ ices.  \otwo\ has a sublimation temperature and pressure comparable to those for CO \citep{Fray:2009} so that the CO models should similarly apply.  The absence of dust in the outbursts observed by Alice suggests a different scenario.

\citet{Skorov:2016}, seeking to explain a narrow, short-lived dust outburst observed by the OSIRIS imager, has proposed a deepening of a pre-existing fracture that would lead to the exposure of a sub-surface ice layer and a subsequent rapid ejection of gas and dust.  Although \citeauthor{Skorov:2016} considered a model with CO ice, their calculations should also be valid for \otwo.  A narrow very short-lived dust jet would be missed by the Alice slit, while the high density of the escaping gas would collisionally be distributed throughout the coma.

\section{SUMMARY}

We report here the detection by the Alice far-ultraviolet spectrograph on {\it Rosetta} of a number of sporadic gas outbursts above the sunward limb.  These outbursts are characterized by sudden increases in the emissions of atomic H and O over a period of 10-30 minutes, without a corresponding enhancement in long wavelength solar reflected light characteristic of dust production.  The emissions are seen to decay over a period of several hours, returning to their quiescent level after a complete rotation of the comet.  Spectroscopic analysis of the ejected gas suggests \otwo\ as the principal driver of the additional gas.  These outbursts do not correlate with any of the visible images of outbursts taken with either OSIRIS or the navigation camera.  A complete accounting of both pre- and post-perihelion events will be presented in a future publication.

\acknowledgments

{\it Rosetta} is an ESA mission with contributions from its member states and NASA.  We thank the members of the {\it Rosetta} Science Ground System and Mission Operations Center teams, in particular Richard Moissl and Michael K\"uppers, for their expert and dedicated help in planning and executing the Alice observations.  We thank Kathleen Mandt and Paul von Allmen for making available their data concerning the May 23 outburst.
The Alice team acknowledges continuing support from NASA's Jet Propulsion Laboratory through contract 1336850 to the Southwest Research Institute.  The work at Johns Hopkins University was supported by a subcontract from Southwest Research Institute.

\vspace{5mm}
\facility{Rosetta}



\listofchanges

\end{document}